\documentclass[prl,showpacs,amsmath,amssymb,superscriptaddress,twocolumn,
floatfix]{revtex4}

\usepackage[english]{babel}
\usepackage{graphicx}
\usepackage{times}
\usepackage{units}

\begin{document}

\title{NRG calculations of the ground-state energy: application to the
correlation effects in the adsorption of magnetic impurities on metal
surfaces}

\author{Rok \v{Z}itko}
\affiliation{J. Stefan Institute, Jamova 39, SI-1000 Ljubljana, Slovenia}

\date{\today}


\begin{abstract}
The ground-state energy of a quantum impurity model can be calculated using
the numerical renormalization group (NRG) with a modified discretization
scheme with sufficient accuracy to reliably extract physical information
about the system. The approach is applied to study binding of magnetic
adsorbates modeled by the Anderson-Newns model for chemisorption on metal
surfaces.
The correlation energy is largest in the valence-fluctuation regime; in the
strong-coupling (Kondo) regime the Kondo-singlet formation energy is found
to be only a minor contribution. As an application of the method to more
difficult surface-science problems, we study the binding energy of a
magnetic atom adsorbed near a step edge on a surface with strongly modulated
surface-state electron density. The zero-temperature magnetic susceptibility
is determined from the field-dependence of the binding energy, thereby
providing an independent result for the Kondo temperature $T_K$, which
agrees very well with the $T_K$ extracted from a thermodynamic calculation. 
\end{abstract}

\maketitle

\newcommand{\expv}[1]{\left\langle #1 \right\rangle}
\newcommand{\korr}[1]{\langle\langle #1 \rangle\rangle}

\renewcommand{\Im}{\mathrm{Im}}
\renewcommand{\Re}{\mathrm{Re}}

\newcommand{\Epsilon}{\mathcal{E}}

\newcommand{\sgn}{\mathrm{sgn}}

The magnetism of nanoscopic objects supported on surfaces is of great
current interest due to possible applications in ultra-dense data storage.
The magnetic properties of adsorbates can now be studied on the single-atom
level using scanning tunneling microscopes (STM) \cite{heinrich2004}. 
Adsorbed atoms attach to metal surfaces by forming strong (covalent) bonds
in a process named chemisorption \cite{norskov1990, brivio1999}. The
chemisorption controls the valence (and thus the magnetic moment) of
magnetic adsorbates, it can lead to adsorbate-induced restructuring of
surfaces, it affects superlattice growth, chemical reactions (catalysis) and
other surface phenomena \cite{brivio1999}. Using an STM, adsorbed atoms may
be manipulated to form artificial nanostructures \cite{crommie1993}. For
successful manipulation of atomic-scale objects it is crucial to understand
the binding properties of adsorbates, i.e. to know the potential-energy
surface as a function of the position of the adsorbate \cite{norskov1990}.

A highly simplified model for studying the chemisorption is the
Anderson-Newns model \cite{anderson1961, newns1969}: $H = H_\mathrm{band} +
H_\mathrm{imp} + H_c$ with
\begin{equation}
\begin{split}
H_\mathrm{band} &= \sum_{k,\sigma\in\{\uparrow,\downarrow\}} \epsilon_k c^\dag_{k\sigma} c_{k\sigma}, \\
H_\mathrm{imp} &= \sum_{\sigma\in\{\uparrow,\downarrow\}} \epsilon n_\sigma + U n_\uparrow n_\downarrow,
\\
H_\mathrm{hyb} &= \sum_{k,\sigma\in\{\uparrow,
\downarrow\}} V_k \left(
c^\dag_{k\sigma} d_\sigma + d^\dag_\sigma c_{k\sigma} \right).
\end{split}
\end{equation}
$H_\mathrm{band}$ describes the continuum of conduction-band electrons with
dispersion $\epsilon_k$, $H_\mathrm{imp}$ corresponds to an adsorbate level
with energy $\epsilon$ and electron-electron repulsion $U$
($n_\sigma=d^\dag_\sigma d_\sigma$ is the level occupancy operator), while
$H_\mathrm{hyb}$ defines the hybridization which can be fully characterized
by the function $\Gamma(\omega)=\sum_{k} |V_k|^2 \delta(\omega-\epsilon_k)$.
The adsorbate binding energy $\Delta E$ is defined as the difference between
the ground state energy of the system described by the full Hamiltonian $H$
and the ground state energy of the decoupled system with $H_\mathrm{hyb}
\equiv 0$ (i.e. the limit where the atom is far away from the surface).
While the Anderson-Newns model was originally intended to describe binding
of hydrogen and alkali atoms, some properties of magnetic adsorbates can
also be studied within a single-orbital approximation \cite{ujsaghy2000}.
General binding properties can be determined qualitatively correctly using
the unrestricted Hartree-Fock method \cite{newns1969}, while the
contributions due to correlations can be calculated variationally
\cite{schonhammer1976}. A method which could very accurately solve the
problem in full generality for arbitrary energy-dependent $\Gamma(\omega)$
and for arbitrarily large interaction strength $U$ has been, however,
lacking. In this work, it is shown that the binding energy can be calculated
with an excellent accuracy using the numerical renormalization group (NRG)
\cite{wilson1975, krishna1980a, bulla2008}.

The NRG consists of a logarithmic discretization of $H_\mathrm{band}$ into
intervals $\left[ \Lambda^{-(j+1)} : \Lambda^{-j} \right]$ with $\Lambda >
1$, followed by a mapping to an effective one-dimensional tight-binding
Hamiltonian with exponentially decreasing hopping constants $\propto
\Lambda^{-i/2}$, and an iterative diagonalisation where one further site is
taken into account at each step. At each iteration $i$, the calculated
excitation spectrum is shifted by subtracting the lowest eigenvalue $E_i$
from all others. The series 
\begin{equation}
E_\mathrm{NRG}=\sum_{i=0}^{\infty} E_i
\end{equation}
is the ground-state energy of the effective Hamiltonian.
To improve the results, several independent NRG calculations are performed
for interleaved discretization meshes shifted by $\Lambda^{1-z}$ with
$z\in(0:1]$ and the final result is obtained as an average over all $z$
\cite{frota1986, campo2005}. To the best knowledge of the author, the
quantity $E_\mathrm{NRG}$ has never been used to extract physical
information about the system, presumably due to poor convergence properties
and systematic errors of the conventional discretization scheme. These
deficiencies of NRG were recently surmounted by a different discretization
approach \cite{resolution, odesolv} which consists of solving the
differential equation
\begin{equation}
\frac{d\Epsilon(x)}{dx} = \frac{\int_{\epsilon(x)}^{\epsilon(x+1)}
\Gamma(\omega) d{\omega}}{\Gamma\left[ \Epsilon(x) \right]},
\end{equation}
where the function $\Epsilon(x)$ with $x=j+z$ yields the discretization
coefficients for each interval $j$ and each parameter $z$; the function
$\epsilon(x)$ defines the discretization grid \cite{odesolv}.

\begin{figure}[htbp]
\includegraphics[width=8cm,clip]{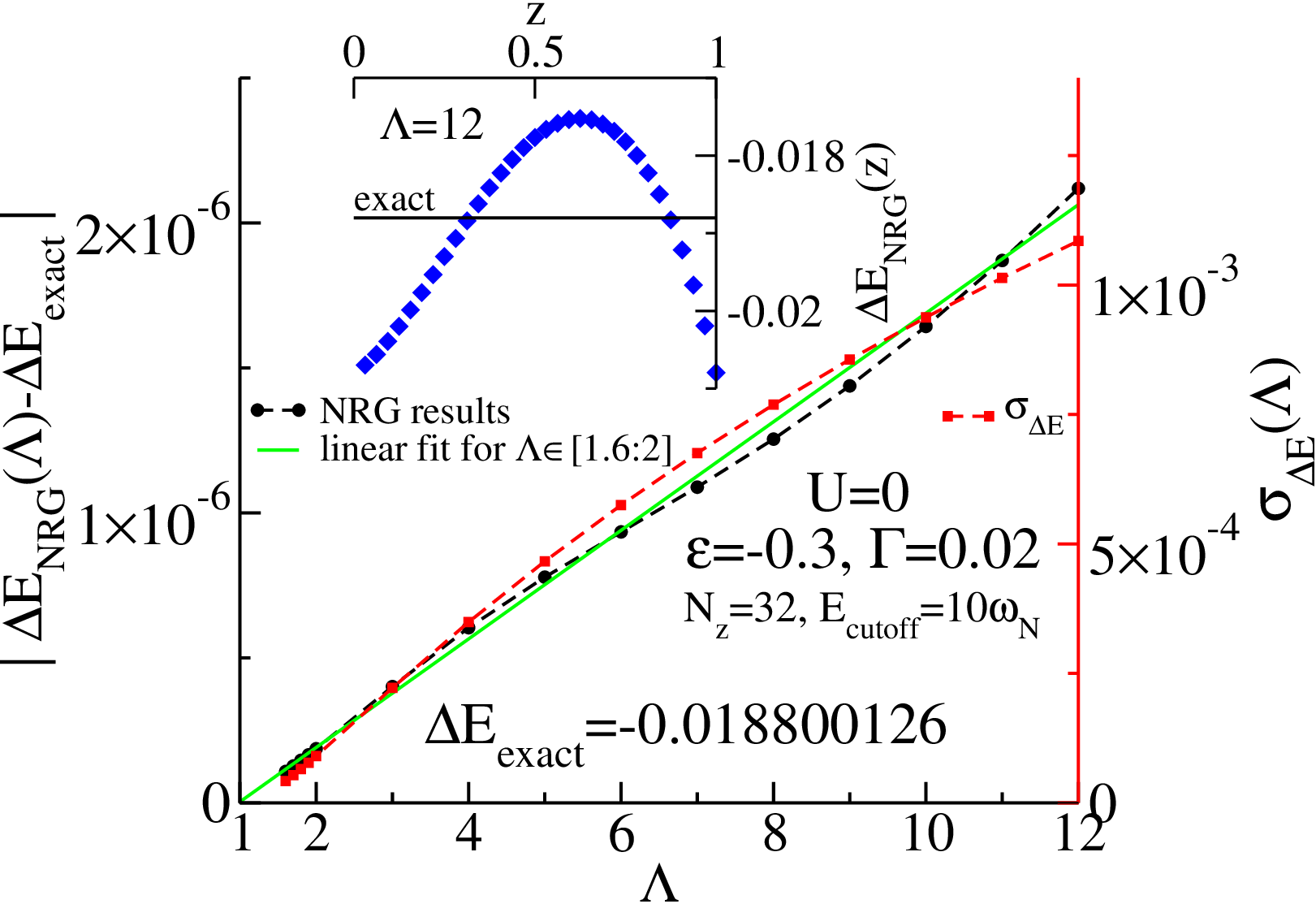}
\caption{(Color online) $\Lambda$-dependence of the calculated binding
energy of a non-interacting adsorbate. Exact binding energy $\Delta
E_\mathrm{exact}$ is subtracted from the numerical results $\Delta
E_\mathrm{NRG}(\Lambda)$ (circles). Full line is a linear fit to results in
the interval $\Lambda\in[1.6:2]$. The error of the extrapolated
$\Lambda\to1$ value is $3.3\times 10^{-9}$. The standard deviation
$\sigma_{\Delta E}$ characterizes the spread of the results for different
parameters $z$. An example of $\Delta E_\mathrm{NRG}(z)$ for $\Lambda=12$ is
shown in the inset. $N_z=32$ different values of $z$ were used, while the
parameter $E_\mathrm{cutoff}=10\omega_N$ defines the truncation cutoff in
the NRG iteration \cite{resolution}.}
\label{fig1}
\end{figure}

We first consider the binding of a non-interacting adsorbate with $U=0$. In
this case, the binding energy can be calculated numerically to arbitrary
precision by a simple quadrature (Eq.~(39) in Ref.~\onlinecite{newns1969}).
For simplicity, we first consider a constant hybridization function:
$\Gamma(\omega) \equiv \Gamma$ for $\omega\in[-1:1]$ and zero otherwise. By
comparing the NRG results with the exact value for a range of discretization
parameters $\Lambda$, Fig.~\ref{fig1}, we find that the binding energies are
calculated with high accuracy even at $\Lambda=12$; for $\Lambda=2$ the
error is $2\times 10^{-7}$. If bare model parameters (bandwidth, $\epsilon$,
$U$) are of the order of the $\unit{eV}$, this magnitude of the error
implies that it is possible to determine the binding energy with $\unit{\mu
eV}$ accuracy. The spread of the results $\Delta E_\mathrm{NRG}(z)$ for
different values of $z$, as measured by the standard deviation
$\sigma_{\Delta E}$ in Fig.~\ref{fig1}, is not an indication of the error
committed but rather contains physically relevant information about the
effects of the hybridization. The $z$-averaging is thus an essential element
of the binding energy calculation and not merely an ad-hoc procedure to
accelerate the convergence.

At large $\Lambda$, the error can be decreased somewhat by increasing $N_z$,
but the improvement is minor. A systematic improvement by one order of the
magnitude can, however, be obtained by interpolation between the data
points, followed by an integration over $z$ on the interval $[0:1]$.  The
error is thereby reduced to $3 \times 10^{-7}$ even at $\Lambda=12$ with no
additional calculations. (There is actually no need to use a uniform mesh of
parameters $z$; it is more economical to choose the $z$-values as the
quadrature nodes.) Using conventional discretization schemes, the errors are
larger by orders of magnitude and even the extrapolated $\Lambda\to1$ value
disagrees with the exact result by $3\times 10^{-4}$; this corresponds to an
error of order $\unit{meV}$, which is barely acceptable especially when
small effects are considered, for example in possible applications to
long-range adsorbate-adsorbate interactions \cite{lau1978}. The use of the
improved discretization scheme from Ref.~\onlinecite{resolution} is thus
crucial and, furthermore, the possibility of obtaining reasonably accurate
results even at large $\Lambda$ implies that calculations can be performed
very efficiently.

\begin{figure}[htbp]
\includegraphics[width=8cm,clip]{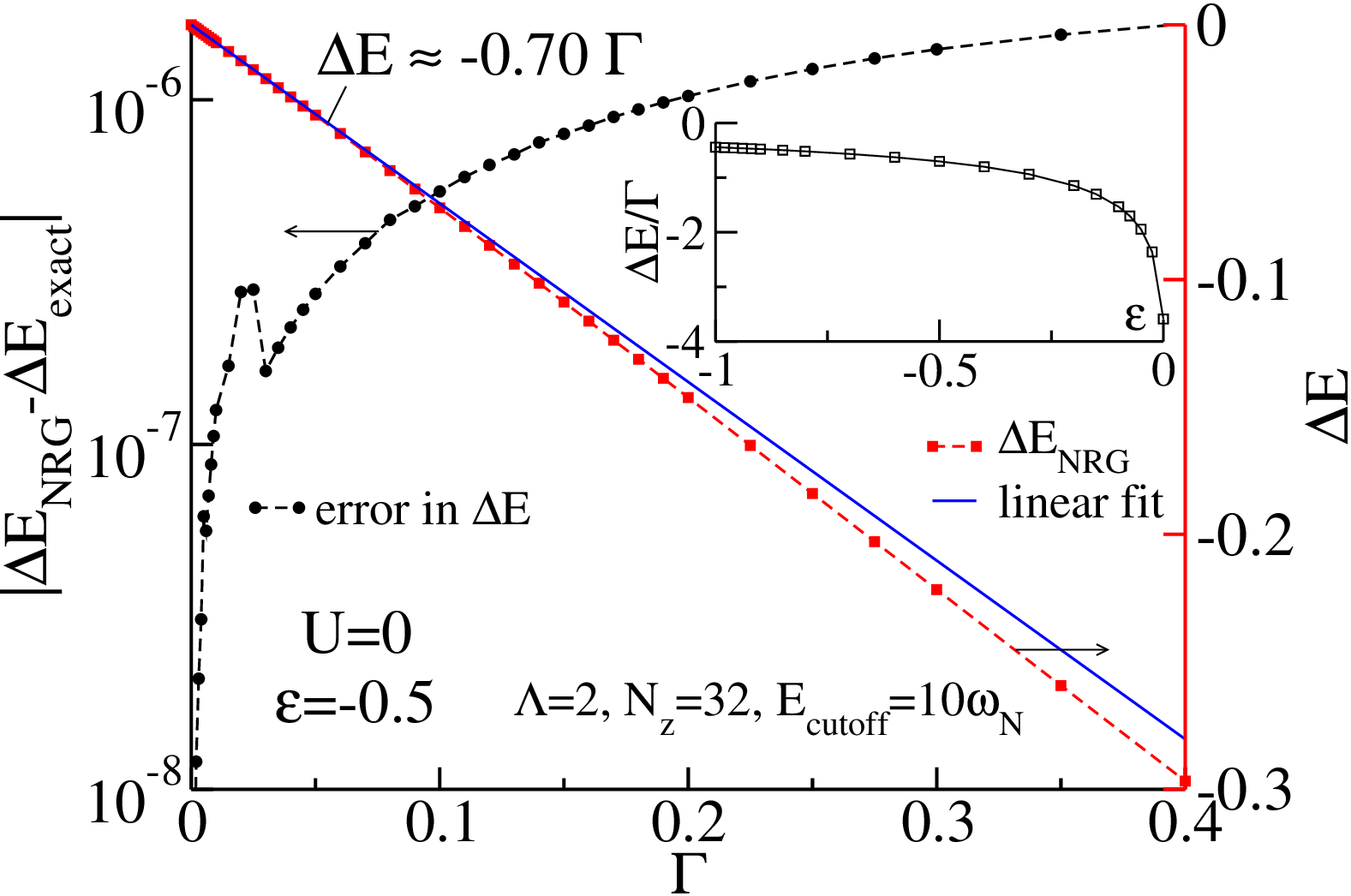}
\caption{(Color online) Binding energy $\Delta E$ (right vertical axis) and
numerical error $\Delta E_\mathrm{NRG}-\Delta E_\mathrm{exact}$ (left
vertical axis) of a non-interacting adsorbate as a function of the
hybridization $\Gamma$. The proportionality coefficient $\Delta
E/\Gamma=-0.70$ is extracted in the interval $\Gamma\in[0:0.01]$. For
reference, the inset shows $\Delta E/\Gamma$ as a function of $\epsilon$.}
\label{fig2}
\end{figure}

For large hybridization $\Gamma$, the adsorbate perturbs the conductance
band more strongly. In NRG calculations, this is of particular concern since
a finite representation of the band is used, thus finite-size effects are
expected to become sizeable. We find, however, that at $\Lambda=2$ the error
is bounded by $1.7\times 10^{-6}$ for all $\Gamma$ in the interval
$[0:0.4]$, Fig.~\ref{fig2}. The binding energy $\Delta E$ is linear in
$\Gamma$ to a good approximation and the coefficient of proportionality
increases in absolute value as $\epsilon$ approaches the Fermi level (see
the inset in Fig.~\ref{fig2}), where the hybridization is more effective in
binding the adsorbate. The adsorbate tends to form a bond with the substrate
by sharing an electron with the conductance-band states. This process is
more efficient when states in the vicinity of the Fermi level are involved,
since their occupancy can be inexpensively changed by the hybridization.
This is similar to bond formation in two-atom molecules, where the binding
energy is largest when the atomic levels are aligned.

\begin{figure}[htbp]
\includegraphics[width=8cm,clip]{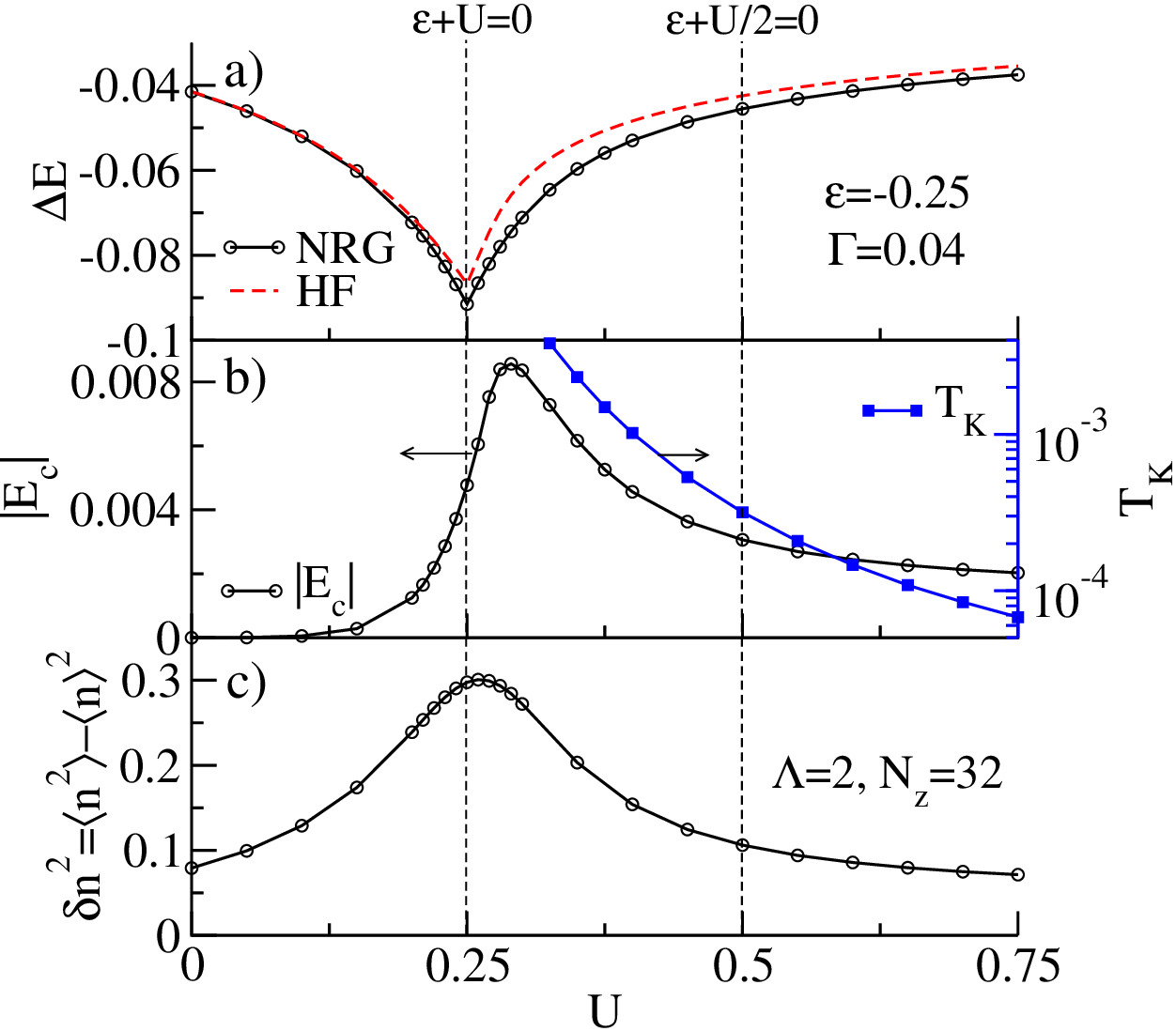}
\caption{(Color online) a) Binding energy, b) correlation energy $E_c=\Delta
E-\Delta E_\mathrm{HF}$, Kondo temperature $T_K$ (right vertical axis), and
c) charge fluctuations of the single-impurity Anderson model as a function
of the electron-electron interaction $U$. The Kondo temperature is extracted
from the thermodynamic properties of the model according to Wilson's
prescription $k_B T_K \chi_\mathrm{imp}(k_B T_K)/(g\mu_B)^2=0.07$, where
$\chi_\mathrm{imp}$ is the impurity magnetic susceptibility
\cite{wilson1975}. }
\label{fig3}
\end{figure}

We now study the full Anderson-Newns model with finite interaction $U$ and
make comparison with the mean-field results obtained using the unrestricted
Hartree-Fock (HF) method (which neglects correlation effects, see also
Ref.~\onlinecite{schonhammer1976}). The binding energy reaches its highest
absolute value for $\epsilon+U=0$ when the single-particle level for an
additional electron crosses the Fermi level, Fig.~\ref{fig3}a. This behavior
is similar to that of the non-interacting model: the binding energy is large
when the charge fluctuates strongly. Both NRG and Hartree-Fock give the same
qualitative features, but it is found that HF underestimates binding. The
additional binding energy can be defined as the ``correlation energy'':
$E_\mathrm{c} = \Delta E-\Delta E_{HF}$. The correlation energy is largest
in the valence fluctuation regime for $\epsilon+U \approx \Gamma$, see
Fig.~\ref{fig3}b. At this point the local moment begins to form (see the
decreasing charge fluctuations $\delta n^2$ in Fig.~\ref{fig3}c for
increasing $U$) and the energy scale of magnetic correlations (the nascent
Kondo regime) is the highest (see the Kondo temperature $T_K$ in
Fig.~\ref{fig3}b). The ``Kondo-singlet formation energy'' of the order of
$T_K$ does not account for the totality of the correlation energy: it is
only a small fraction, in particular in the large-$U$ limit where the Kondo
temperature is strongly suppressed. The most important contribution to the
correlation energy thus stems from local charge correlations, rather than
from extended Kondo correlations.

\begin{figure}[htbp]
\includegraphics[width=8cm,clip]{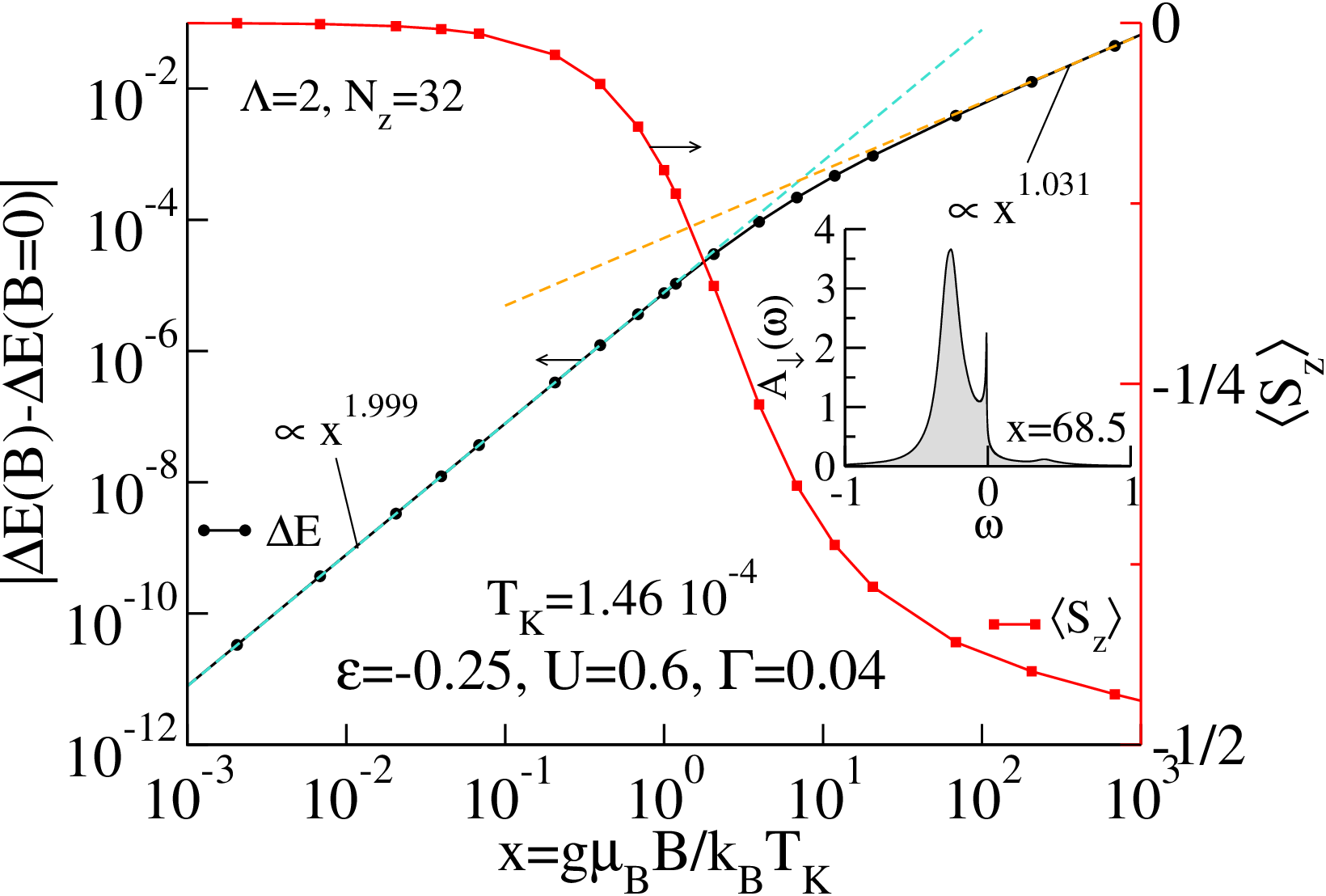}
\caption{(Color online) Binding energy and spin polarization of a magnetic
adsorbate in an external magnetic field (expressed in reduced units of
$x=g\mu_B B/k_B T_K$). The inset shows the spin-resolved impurity spectral
function in a strong field. }
\label{fig4}
\end{figure}

The energy gain due to Kondo correlations is lost in a strong magnetic
field, see Fig.~\ref{fig4}. The quadratic reduction for low fields ($g\mu_B
B \ll k_B T_K$) is expected due to finite spin susceptibility at zero
temperature in the strong-coupling regime \cite{wilson1975}. From the
prefactor we can extract the zero-temperature magnetic susceptibility
\begin{equation}
\chi(T=0) = \frac{W (g\mu_B)^2}{4\pi k_B T_K},
\end{equation}
where $W\approx 1.29026$ is the Wilson number \cite{wilson1975, andrei1983}.
From $\chi(T=0)$ we then obtain a value $T_K'=1.43\times 10^{-4}$ for the
Kondo temperature, which differs from the value of $T_K=1.46\times 10^{-4}$
determined in a thermodynamic calculation of magnetic susceptibility by
$<3\%$. Considering that the values are obtained using entirely independent
procedures, their close agreement is an exceptional confirmation of the
method. The remaining small discrepancy stems mostly from the error
associated with obtaining the coefficient of the $B^2$ contribution to the
total energy in the limit of very small magnetic fields. For large fields
($g \mu_B B \gg k_B T_K$), the Zeeman effect takes over and the variation of
$\Delta E$ is approximately linear.

\begin{figure}[htbp]
\includegraphics[width=8cm,clip]{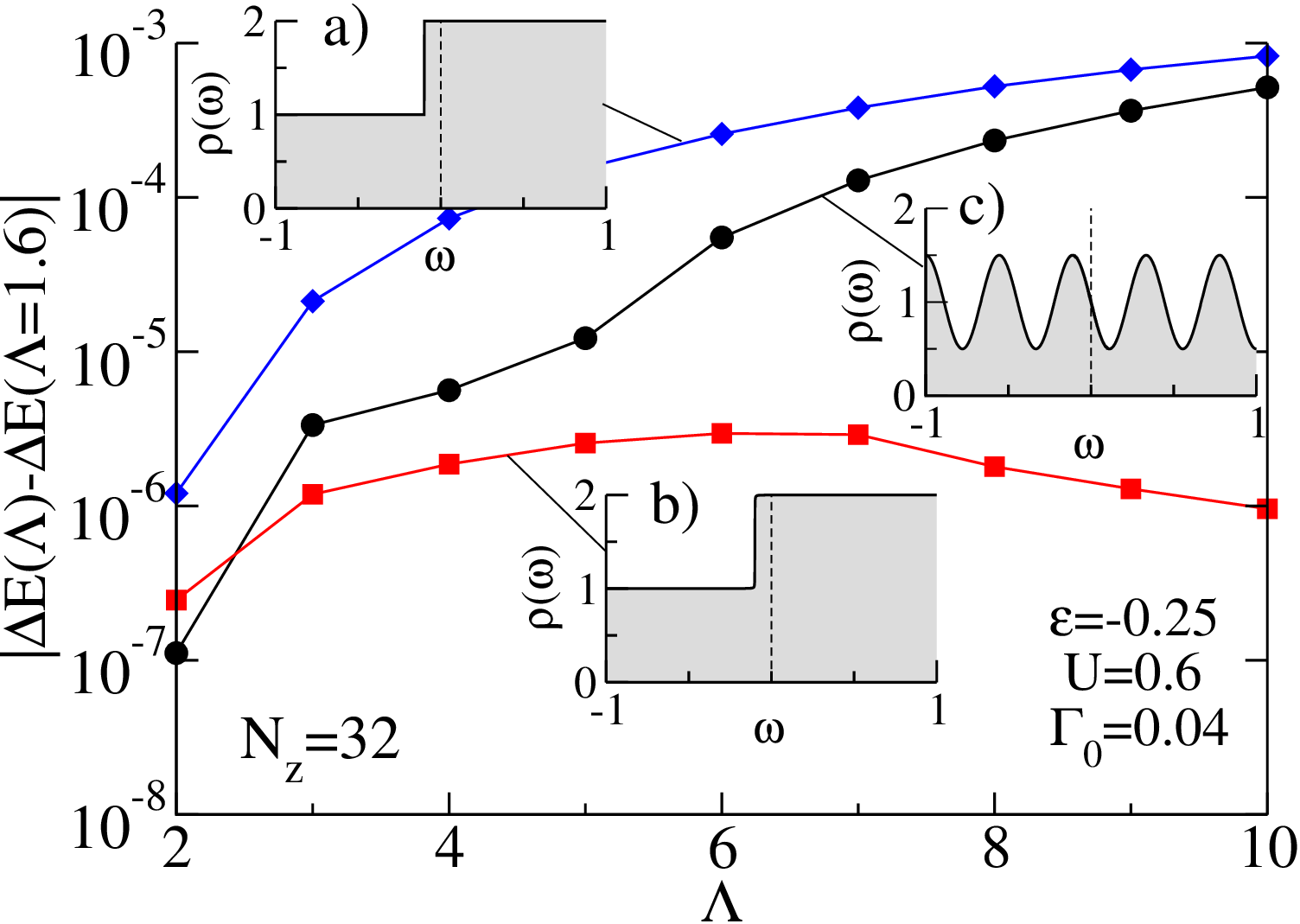}
\caption{(Color online) $\Lambda$-dependence of the binding energy of a
magnetic adsorbate hybridized to a band with energy-dependent density of
states $\rho(\omega)$. The hybridization function is
$\Gamma(\omega)=\Gamma_0 \rho(\omega)$ with a) sharp step function
$\rho(\omega)=1+\theta(\omega-\omega_0)$ with $\omega_0=-0.1$, b) rounded
step-function $\rho(\omega)= 1 + (1/2+ (1/\pi) \tan^{-1} \left[
\pi(\omega-\omega_0)/\Delta \right] )$, where $\Delta=0.001$, and c)
oscillatory $\rho(\omega)=1+(1/2) \cos[(9/2)\pi(1+\omega)]$. The $\Lambda=1.6$
results are used as reference values and subtracted from $\Delta
E(\Lambda)$. }
\label{fig5}
\end{figure}

Albeit constant hybridization is a convenient simplification, in realistic
problems $\Gamma(\omega)$ is strongly energy dependent. Three forms are
considered here: sharp and rounded step functions, and an oscillatory
function. The convergence with $\Lambda$ depends significantly on the form,
see Fig.~\ref{fig5}: while the error remains approximately constant at
$\approx 10^{-6}$ for the rounded step function, it increases significantly
for the sharp-step and oscillatory function. As expected, sharp
discontinuities and variations that occur over extended energy intervals
lead to larger errors than smooth localized changes.

\begin{figure}[htbp]
\includegraphics[width=8cm,clip]{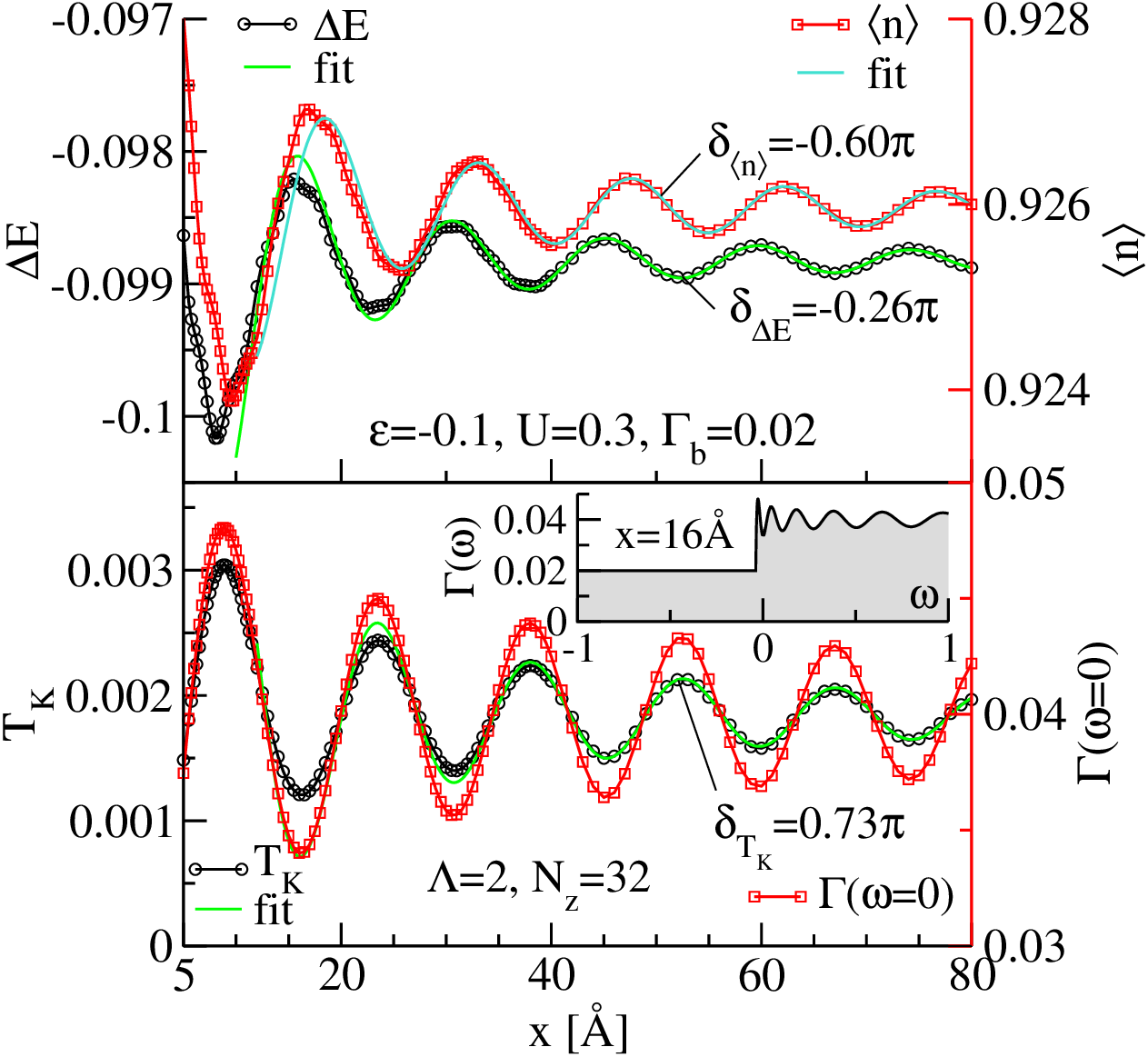}
\caption{(Color online) Properties of a magnetic adsorbate on a surface at a
distance of $x$ from a step edge modelled as a hard-wall potential
scatterer. Fitting by oscillatory power-law functions $A+B \cos(2k_F
x+\delta)/x^\alpha$ provides the phase shifts $\delta$ (show in the figure)
and decay constants $\alpha$: the binding energy $\Delta E$ decays as
$1/x^{3/2}$, the occupancy as $1/x^{1.19}$, and the Kondo temperature as
$1/x^{1.16}$. The fitting procedure was performed with the data in the
asymptotic region $x \in [40:80]\unit{\AA}$. The inset shows the
hybridization function at $x=\unit[16]{\AA}$. $\omega_0=-0.039$,
$k_F=\unit[0.217]{\AA^{-1}}$.}
\label{fig6}
\end{figure}

The capabilities of the method for problems with strongly energy-dependent
hybridization are demonstrated with the example of a magnetic adsorbate in
the vicinity of a step edge on a surface supporting a surface-state band.
The adsorbate hybridizes both with the bulk states via $\Gamma_b$ (which
will be assumed not to vary with energy) and with surface states via
$\Gamma_s$. On a clean flat surface, $\Gamma(\omega) = \Gamma_b + \Gamma_s
\theta(\omega-\omega_0)$, where $\omega_0$ is the onset of the surface-state
band. More interesting situation occurs when the adsorbate is adsorbed near
a step edge, where the local density of states of surface-state electrons is
modulated by standing waves. Modelling the step edge as a hard-wall
potential, the energy-resolved charge density 
is $\delta n(x,\omega) \propto 1-J_0[2k(\omega) x]$, where $J_0$ is the
Bessel function, $k(\omega)$ the wave-number at energy $\omega$ and $x$ the
distance from the step edge. Modelling the surface-state electrons as free
electrons with effective mass $m^*$ we have
$k(\omega)=\left[(2m^*/\hbar) \left( \omega-\omega_0
\right)\right]^{1/2}$.  
The hybridization function is thus
\begin{equation}
\Gamma(\omega)= \Gamma_b + \Gamma_s \theta\left(\omega-\omega_0 \right) 
\left\{ 1 - J_0 \left[2k(\omega)x\right] \right\}.
\end{equation}
While it is by now established that for magnetic impurities on noble-metal
surfaces $\Gamma_s \ll \Gamma_b$ \cite{lin2006, henzl2007}, we will
nevertheless take a greatly exaggerated ratio $\Gamma_s/\Gamma_b=1$ to
accentuate the effect of the energy-dependence of $\Gamma(\omega)$. In fact,
on surfaces with giant Friedel oscillations \cite{sprunger1997} such ratio
might be realistic.

The adsorbate properties reflect the oscillatory features in
$\Gamma(\omega)$, see Fig.~\ref{fig6}. The Kondo temperature is strongly
correlated with the variation of $\Gamma$ at the Fermi level and it can be
well described by a cosine function with constant phase shift $\delta_{T_K}$
multiplied by some envelope function which is, to a good approximation, a
power-law decay $1/x^{1.16}$. The binding energy, however, exhibits some
additional structure, in particular for low values of $x$. (This is not a
numerical artefact: the same result is obtained for other choices of NRG
parameters.) Qualitatively similar features are visible in the adsorbate
level occupancy $\expv{n}$, but at shifted positions $x$. The origin of
these effects is thus in the details of the energy dependence of
$\Gamma(\omega)$ over a wide energy interval (i.e. on the atomic scale of
$\epsilon$ and $U$). This is unlike the Kondo temperature, which depends
mostly on the values of $\Gamma(\omega)$ in the narrow interval on the scale
of $T_K$ itself and therefore simply follows the variation of
$\Gamma(\omega=0)$. It may be noted that strong binding corresponds to high
Kondo temperature and that variations of $\Delta E$ and $T_K$ are of the
same order of magnitude, pointing to a large effect of magnetic correlation
in this situation.

The NRG method is a very capable tool for studying correlation effects in
magnetic adsorbates on surfaces. The demonstrated favorable scaling of
errors with $\Lambda$ brings more realistic (multi-orbital) models within
the reach of modern computing facilities. The technique for calculating
ground-state energies is very general and it can be, for example, applied to
calculate the response of the system (expectation values, susceptibilities)
with respect to arbitrary perturbations.

\bibliography{paper}

\begin{thebibliography}{20}
\expandafter\ifx\csname natexlab\endcsname\relax\def\natexlab#1{#1}\fi
\expandafter\ifx\csname bibnamefont\endcsname\relax
  \def\bibnamefont#1{#1}\fi
\expandafter\ifx\csname bibfnamefont\endcsname\relax
  \def\bibfnamefont#1{#1}\fi
\expandafter\ifx\csname citenamefont\endcsname\relax
  \def\citenamefont#1{#1}\fi
\expandafter\ifx\csname url\endcsname\relax
  \def\url#1{\texttt{#1}}\fi
\expandafter\ifx\csname urlprefix\endcsname\relax\def\urlprefix{URL }\fi
\providecommand{\bibinfo}[2]{#2}
\providecommand{\eprint}[2][]{\url{#2}}

\bibitem[{\citenamefont{Heinrich et~al.}(2004)\citenamefont{Heinrich, Gupta,
  Lutz, and Eigler}}]{heinrich2004}
\bibinfo{author}{\bibfnamefont{A.~J.} \bibnamefont{Heinrich}},
  \bibinfo{author}{\bibfnamefont{J.~A.} \bibnamefont{Gupta}},
  \bibinfo{author}{\bibfnamefont{C.~P.} \bibnamefont{Lutz}}, \bibnamefont{and}
  \bibinfo{author}{\bibfnamefont{D.~M.} \bibnamefont{Eigler}},
  \bibinfo{journal}{Science} \textbf{\bibinfo{volume}{306}},
  \bibinfo{pages}{466} (\bibinfo{year}{2004}).

\bibitem[{\citenamefont{N{\o}rskov}(1990)}]{norskov1990}
\bibinfo{author}{\bibfnamefont{J.~K.} \bibnamefont{N{\o}rskov}},
  \bibinfo{journal}{Rep. Prog. Phys.} \textbf{\bibinfo{volume}{53}},
  \bibinfo{pages}{1253} (\bibinfo{year}{1990}).

\bibitem[{\citenamefont{Brivio and Trioni}(1999)}]{brivio1999}
\bibinfo{author}{\bibfnamefont{G.~P.} \bibnamefont{Brivio}} \bibnamefont{and}
  \bibinfo{author}{\bibfnamefont{M.~I.} \bibnamefont{Trioni}},
  \bibinfo{journal}{Rev. Mod. Phys.} \textbf{\bibinfo{volume}{71}},
  \bibinfo{pages}{231} (\bibinfo{year}{1999}).

\bibitem[{\citenamefont{Crommie et~al.}(1993)\citenamefont{Crommie, Lutz, and
  Eigler}}]{crommie1993}
\bibinfo{author}{\bibfnamefont{M.~F.} \bibnamefont{Crommie}},
  \bibinfo{author}{\bibfnamefont{C.~P.} \bibnamefont{Lutz}}, \bibnamefont{and}
  \bibinfo{author}{\bibfnamefont{D.~M.} \bibnamefont{Eigler}},
  \bibinfo{journal}{Science} \textbf{\bibinfo{volume}{262}},
  \bibinfo{pages}{218} (\bibinfo{year}{1993}).

\bibitem[{\citenamefont{Anderson}(1961)}]{anderson1961}
\bibinfo{author}{\bibfnamefont{P.~W.} \bibnamefont{Anderson}},
  \bibinfo{journal}{Phys. Rev.} \textbf{\bibinfo{volume}{124}},
  \bibinfo{pages}{41} (\bibinfo{year}{1961}).

\bibitem[{\citenamefont{Newns}(1969)}]{newns1969}
\bibinfo{author}{\bibfnamefont{D.~M.} \bibnamefont{Newns}},
  \bibinfo{journal}{Phys. Rev.} \textbf{\bibinfo{volume}{178}},
  \bibinfo{pages}{1123} (\bibinfo{year}{1969}).

\bibitem[{\citenamefont{Ujsaghy et~al.}(2000)\citenamefont{Ujsaghy, Kroha,
  Szunyogh, and Zawadowski}}]{ujsaghy2000}
\bibinfo{author}{\bibfnamefont{O.}~\bibnamefont{Ujsaghy}},
  \bibinfo{author}{\bibfnamefont{J.}~\bibnamefont{Kroha}},
  \bibinfo{author}{\bibfnamefont{L.}~\bibnamefont{Szunyogh}}, \bibnamefont{and}
  \bibinfo{author}{\bibfnamefont{A.}~\bibnamefont{Zawadowski}},
  \bibinfo{journal}{Phys. Rev. Lett.} \textbf{\bibinfo{volume}{85}},
  \bibinfo{pages}{2557} (\bibinfo{year}{2000}).

\bibitem[{\citenamefont{Sch\"onhammer}(1976)}]{schonhammer1976}
\bibinfo{author}{\bibfnamefont{K.}~\bibnamefont{Sch\"onhammer}},
  \bibinfo{journal}{Phys. Rev. B} \textbf{\bibinfo{volume}{13}},
  \bibinfo{pages}{4336} (\bibinfo{year}{1976}).

\bibitem[{\citenamefont{Wilson}(1975)}]{wilson1975}
\bibinfo{author}{\bibfnamefont{K.~G.} \bibnamefont{Wilson}},
  \bibinfo{journal}{Rev. Mod. Phys.} \textbf{\bibinfo{volume}{47}},
  \bibinfo{pages}{773} (\bibinfo{year}{1975}).

\bibitem[{\citenamefont{Krishna-murthy
  et~al.}(1980)\citenamefont{Krishna-murthy, Wilkins, and
  Wilson}}]{krishna1980a}
\bibinfo{author}{\bibfnamefont{H.~R.} \bibnamefont{Krishna-murthy}},
  \bibinfo{author}{\bibfnamefont{J.~W.} \bibnamefont{Wilkins}},
  \bibnamefont{and} \bibinfo{author}{\bibfnamefont{K.~G.}
  \bibnamefont{Wilson}}, \bibinfo{journal}{Phys. Rev. B}
  \textbf{\bibinfo{volume}{21}}, \bibinfo{pages}{1003} (\bibinfo{year}{1980}).

\bibitem[{\citenamefont{Bulla et~al.}(2008)\citenamefont{Bulla, Costi, and
  Pruschke}}]{bulla2008}
\bibinfo{author}{\bibfnamefont{R.}~\bibnamefont{Bulla}},
  \bibinfo{author}{\bibfnamefont{T.}~\bibnamefont{Costi}}, \bibnamefont{and}
  \bibinfo{author}{\bibfnamefont{T.}~\bibnamefont{Pruschke}},
  \bibinfo{journal}{Rev. Mod. Phys.} \textbf{\bibinfo{volume}{80}},
  \bibinfo{pages}{395} (\bibinfo{year}{2008}).

\bibitem[{\citenamefont{Frota and Oliveira}(1986)}]{frota1986}
\bibinfo{author}{\bibfnamefont{H.~O.} \bibnamefont{Frota}} \bibnamefont{and}
  \bibinfo{author}{\bibfnamefont{L.~N.} \bibnamefont{Oliveira}},
  \bibinfo{journal}{Phys. Rev. B} \textbf{\bibinfo{volume}{33}},
  \bibinfo{pages}{7871} (\bibinfo{year}{1986}).

\bibitem[{\citenamefont{Campo and Oliveira}(2005)}]{campo2005}
\bibinfo{author}{\bibfnamefont{V.~L.} \bibnamefont{Campo}} \bibnamefont{and}
  \bibinfo{author}{\bibfnamefont{L.~N.} \bibnamefont{Oliveira}},
  \bibinfo{journal}{Phys. Rev. B} \textbf{\bibinfo{volume}{72}},
  \bibinfo{pages}{104432} (\bibinfo{year}{2005}).

\bibitem[{\citenamefont{\v{Z}itko and Pruschke}(2009)}]{resolution}
\bibinfo{author}{\bibfnamefont{R.}~\bibnamefont{\v{Z}itko}} \bibnamefont{and}
  \bibinfo{author}{\bibfnamefont{T.}~\bibnamefont{Pruschke}},
  \bibinfo{journal}{Phys. Rev. B} \textbf{\bibinfo{volume}{79}},
  \bibinfo{pages}{085106} (\bibinfo{year}{2009}).

\bibitem[{\citenamefont{\v{Z}itko}(2009)}]{odesolv}
\bibinfo{author}{\bibfnamefont{R.}~\bibnamefont{\v{Z}itko}},
  \bibinfo{journal}{Comp. Phys. Comm.}  (\bibinfo{year}{2009}).

\bibitem[{\citenamefont{Lau and Kohn}(1978)}]{lau1978}
\bibinfo{author}{\bibfnamefont{K.~H.} \bibnamefont{Lau}} \bibnamefont{and}
  \bibinfo{author}{\bibfnamefont{W.}~\bibnamefont{Kohn}},
  \bibinfo{journal}{Surf. Sci.} \textbf{\bibinfo{volume}{75}},
  \bibinfo{pages}{69} (\bibinfo{year}{1978}).

\bibitem[{\citenamefont{Andrei et~al.}(1983)\citenamefont{Andrei, Furuya, and
  Lowenstein}}]{andrei1983}
\bibinfo{author}{\bibfnamefont{N.}~\bibnamefont{Andrei}},
  \bibinfo{author}{\bibfnamefont{K.}~\bibnamefont{Furuya}}, \bibnamefont{and}
  \bibinfo{author}{\bibfnamefont{J.~H.} \bibnamefont{Lowenstein}},
  \bibinfo{journal}{Rev. Mod. Phys.} \textbf{\bibinfo{volume}{55}},
  \bibinfo{pages}{331} (\bibinfo{year}{1983}).

\bibitem[{\citenamefont{Lin et~al.}(2006)\citenamefont{Lin, Neto, and
  Jones}}]{lin2006}
\bibinfo{author}{\bibfnamefont{C.-Y.} \bibnamefont{Lin}},
  \bibinfo{author}{\bibfnamefont{A.~H.~C.} \bibnamefont{Neto}},
  \bibnamefont{and} \bibinfo{author}{\bibfnamefont{B.~A.} \bibnamefont{Jones}},
  \bibinfo{journal}{Phys. Rev. Lett.} \textbf{\bibinfo{volume}{97}},
  \bibinfo{pages}{156102} (\bibinfo{year}{2006}).

\bibitem[{\citenamefont{Henzl and Morgenstern}(2007)}]{henzl2007}
\bibinfo{author}{\bibfnamefont{J.}~\bibnamefont{Henzl}} \bibnamefont{and}
  \bibinfo{author}{\bibfnamefont{K.}~\bibnamefont{Morgenstern}},
  \bibinfo{journal}{Phys. Rev. Lett.} \textbf{\bibinfo{volume}{98}},
  \bibinfo{pages}{266601} (\bibinfo{year}{2007}).

\bibitem[{\citenamefont{Sprunger et~al.}(1997)\citenamefont{Sprunger, Petersen,
  Plummer, Laegsgaard, and Besenbacher}}]{sprunger1997}
\bibinfo{author}{\bibfnamefont{P.~T.} \bibnamefont{Sprunger}},
  \bibinfo{author}{\bibfnamefont{L.}~\bibnamefont{Petersen}},
  \bibinfo{author}{\bibfnamefont{E.~W.} \bibnamefont{Plummer}},
  \bibinfo{author}{\bibfnamefont{E.}~\bibnamefont{Laegsgaard}},
  \bibnamefont{and}
  \bibinfo{author}{\bibfnamefont{F.}~\bibnamefont{Besenbacher}},
  \bibinfo{journal}{Science} \textbf{\bibinfo{volume}{275}},
  \bibinfo{pages}{1764} (\bibinfo{year}{1997}).

\end{thebibliography}

\end{document}